\documentclass[trackchanges]{aastex701}

\usepackage{makecell}
\usepackage{amsmath}

\begin{document}


\title{Revisiting the Rheology of Neutron Star Crusts with Molecular Dynamics}

\author[0009-0006-2122-5606]{Matthew E. Caplan}
\email{Corresponding author: mecapl1@ilstu.edu}
\affiliation{Department of Physics, Illinois State University,
Normal, IL 61761, USA}
\affiliation{Department of Physics, University of Illinois Urbana–Champaign, Urbana, IL 61801, US}

\author[0000-0002-9711-9424]{Ashley Bransgrove} %
\email{abransgrove@princeton.edu}
\affil{Princeton Center for Theoretical Science and Department of Astrophysical Sciences, Princeton University, Princeton, NJ 08544, USA}
\affil{Physics Department and Columbia Astrophysics Laboratory, Columbia University, 538 West 120th Street, New York, NY 10027}

\begin{abstract}
Explosive events from magnetars are likely due to the catastrophic release of stress in their crusts, but the behavior of crustal matter beyond linear elasticity is poorly understood. We argue here that seminal results from molecular dynamics informing crust breaking calculations are non-converged, and must be revisited. We estimate the criteria for quasi-static, rate-independent flow by comparing imposed deformation timescales to grain boundary diffusion in polycrystals. We argue that convergence in this regime should be observed at strain rates slower than $10^{-5}\,\omega_p$ (plasma frequency $\omega_p$) in simulations of $N\approx10^5$ particles across order 10 grains at a quarter of the melting temperature. Though computationally expensive, this is tractable with modern methods and GPU supercomputers.
\end{abstract}




\section{Introduction}

The dynamics of neutron star crusts in the linear elastic regime is well understood. However, when the crustal matter is stressed beyond its critical strain $\varepsilon \approx  0.1$ mechanical failure occurs. Neutron star crusts cannot break like terrestrial materials because the formation of voids that facilitate crack propagation is suppressed at high pressure. Therefore, neutron star crust breaking more likely involves plastic flow. 

The seminal work of \citet[hereafter HK09]{HorowitzKadau} observes highly unusual flow. They report on molecular dynamics (MD) of a deformed polycrystal and observe strain softening where stress decreases with increasing strain beyond the yield stress. This result informs much of the literature. 
Given the computational limitations of the time, HK09 only report on one simulation of a polycrystal, and do not test whether the post-break behavior is independent of the imposed strain rate. Generally, one must verify for convergence that the deformation is sufficiently slow to be quasi-static. 
We argue below that the simulation in HK09 is not strained sufficiently slowly to be converged.

In a Coulomb crystal, the natural microscopic timescale is set by the ion plasma frequency ${\omega_p = \sqrt{ 4 \pi e^2 Z^2 n_i/A m_n}}$, with $eZ$ and $Am_n$ the nuclear charge and mass, and $n_i$ the ion number density. 
A macroscopic timescale comes from elastic signal propagation across the simulation domain. The elastic wave speed is $v_{\rm el} = \sqrt{\mu/ \rho}$, with $\mu$ the shear modulus and $\rho = n_i A m_n$ the mass density. For a Coulomb solid, $\mu \simeq 0.11 \, n_i (Z e)^2 /a_i$ and $a_i = \left(3/4\pi n_i\right)^{1/3}$ the Wigner-Seitz radius.

If the MD volume is deformed (for example, from a cube to a parallepiped) at a strain rate $\dot{\epsilon}$, the characteristic velocity associated with the box deformation is $v_{\mathrm{box}} = \dot{\epsilon} L$
where $L = (N/n_i)^{1/3}$ is the cubic box size containing $N$ nuclei. For quasi-static deformation the elastic information must cross the simulation domain on a timescale much shorter than the imposed deformation, \textit{i.e.}, $v_{\mathrm{box}}/v_{\rm el} \ll 1.$ Combining expressions, 
\begin{equation}
\frac{v_{\mathrm{box}}}{v_{\rm el}} 
= \dot{\epsilon} N^{1/3} \omega_p^{-1}
\frac{\sqrt{4\pi}}{\sqrt{0.11}}\left(\frac{3}{4\pi}\right)^{1/6}
\simeq 8.5 \, N^{1/3} \, \frac{\dot{\epsilon}}{\omega_p}.
\end{equation}

\noindent The parameters reported in \cite{HorowitzKadau} are $N \simeq 1.28\times 10^{7}$ and $\dot{\epsilon} = 4\times 10^{-7} \, c/\mathrm{fm}$ with $\omega_p \simeq 3.7\times 10^{-3} \, c/\mathrm{fm}$ (from $Z=29.4, \, A=88, \,n_i = 7.18\times10^{-5} \ \rm{fm}^{-3}$), giving ${\dot{\epsilon}/\omega_p \simeq 1.1\times 10^{-4}}$. Therefore, $v_{\mathrm{box}}/v_{\rm el} = 0.21$. 
This suggests that the deformation is barely quasi-static and that main results of HK09, namely strain softening and amorphization, are likely a result of strain rate dependence in the near shocked limit. To revisit HK09 and replicate their simulation at slower strain rates would require thousands of node hours. We check here if this is feasible by estimating the minimum $N$ and $\dot\epsilon$ for convergence. 

\section{Convergence Criteria}

Nuclei have the greatest mobility at grain boundaries (GBs), where atomic packing is less ordered and diffusion rates are significantly enhanced \citep{Hughto2011Diffusion}. 
Maximizing the volume filling fraction of GBs will maximize the number density of dislocations and volume where slip occurs, and is equivalent to minimizing grain size. 
Consider a single quasi-spherical grain of radius $d$ that is being slowly sheared. The nucleation of a thin film of super-cooled liquid (GB) inside the grain would allow slippage of the two grain halves that would reduce the strain. 
However, as the preferred phase at these densities/temperatures is a body centered cubic crystal, energy input is required to nucleate the GB. This energy can be supplied by elastic energy at the critical strain $\epsilon_c\sim 0.1$. This gives the typical strain energy required to nucleate a GB 
\begin{equation}
    f \mu \pi d^2 \delta \sim \frac{1}{2}\mu \epsilon_c ^2 \frac{4 \pi}{3}d^3,
    \label{energy}
\end{equation}
where $\delta \sim  2a_i$ is the thickness of the grain boundary, while $f$ is a small factor that represents the energy difference between the crystal and liquid states\footnote{The energy of the strained grain should equal the energy required to nucleate a GB plus the energy dissipated by sliding. However the energy dissipated by sliding is comparable to the nucleation energy, and therefore only contributes a factor $\sim2$ to the left side of Eq.~\ref{energy}.}. We take $f\sim 10^{-2}$ in units of $e^2 Z^2/a_i$; Madelung energies and potential energies per nuclei for random solids are available in the literature, these assume infinite media. 
This gives the preferred grain size at the critical strain 
\begin{equation}
    d \sim 2f\frac{\delta}{\epsilon_c ^2} 
    \sim 10 \, a_i 
\end{equation}
Grains larger than this are likely to break because they have sufficient free energy to nucleate GBs. 
For the typical dimensionless screening $\kappa = a_i / \lambda\approx 0.66$, MD simulations demonstrate convergence for Coulomb interaction cut-off distances $\gtrsim 10 a_i$. This suggests that for grains larger than $\sim 10 a_i$ boundaries on opposite sides of the grain do not strongly influence each other. We argue that a grain size of order $10-20 a_i$ is ideal, or $20^3 \approx 10^4$ particles per grain is a reasonable minimum. Assuming a minimum of 10 grains in the MD simulation suggests that no fewer than $10^5$ nuclei are required. GBs being about $\delta=2a_i$ thick achieves a volume filling fraction of boundaries of order $f_{\rm GB} \equiv \delta/d \sim0.1$.

\begin{figure}
    \centering
    \includegraphics[trim=25 115 195 20, clip,width=0.50\linewidth]{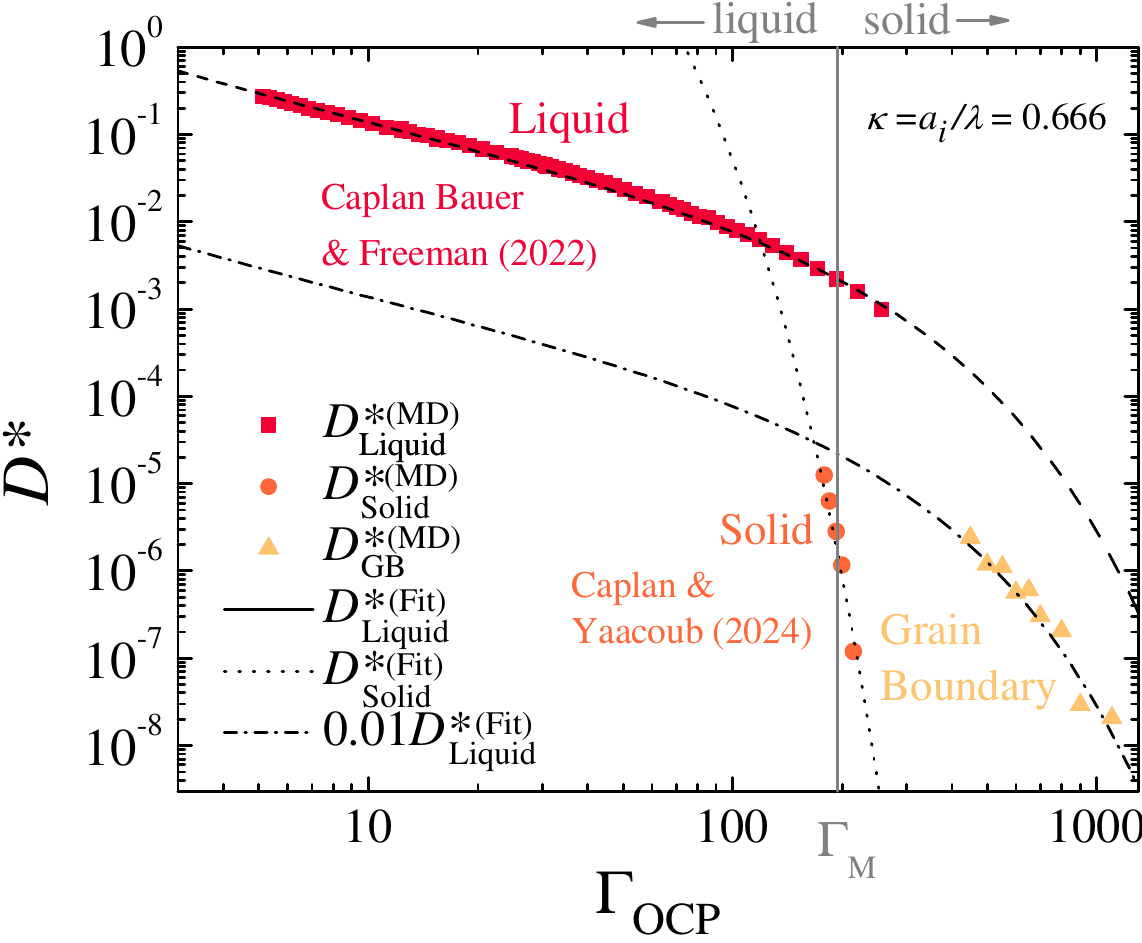}
    \caption{Dimensionless diffusion coefficients $D^* = D/\omega_p a_i^2$ for Yukawa plasmas at a typical screenings for neutron star crusts. Points that cross melting are from weakly superheated (supercooled) simulations.}
    \label{fig:Dall}
\end{figure}

We now estimate the required $\dot\epsilon$ to achieve quasi-static rate independent flow. Internal relaxation mechanisms must remain fast compared to the imposed strain rate. If the dominant relaxation mechanism in a polycrystalline Coulomb solid is grain boundary mobility, the characteristic diffusion time is $\tau_{\rm GB} \approx \ell^2/D_{\rm GB}$, where $\ell$ is the relevant transport length scale, taken here to be the grain boundary thickness $\ell \sim 2a_i$. For a polycrystal with grain size $d$, the grain boundary volume filling fraction is again $f_{\rm GB} \approx (2a_i)/d$, so the effective relation is reduced to $D_{\rm eff} \sim f_{\rm GB} D_{\rm GB}$. Quasi-static flow therefore requires $\dot{\epsilon} \lesssim D_{\rm eff}/\ell^2$, or equivalently $\dot{\epsilon} \lesssim f_{\rm GB} D_{\rm GB}/(2a_i)^2$.

Diffusion coefficients depend strongly on the phase (solid or liquid), shown in Fig. \ref{fig:Dall}. In the strongly coupled limit ($\Gamma_{\rm M} \gtrsim175$ where $\Gamma = e^2 Z^2/a_ikT$), ${D^* \propto \exp(-B\Gamma)}$ where $B_\mathrm{liquid}=0.006$ (dashed) and $B_\mathrm{solid}=0.103$ (dotted) from the fits obtained in \citet{CaplanBauerFreeman} and \citet{CaplanYaacoub2025}. For realistic neutron star crust temperatures, perhaps $T = 0.25 T_\mathrm{M}$ (melting temperature $T_{\rm M}$, or $\Gamma=\Gamma_{\rm M}/0.25$), diffusion rates in a perfect solid are more than 10 orders of magnitude smaller than that of a supercooled liquid. New MD simulations with GBs from \cite{GBD2025} find that GBs, being amorphous, have mobility comparable to a supercooled liquid suppressed by two orders of magnitude.

For $T \sim 0.25 T_{\rm M}$, Fig. \ref{fig:Dall} indicates $D_{\rm GB} \approx 10^{-7} \, \omega_p a_i^2$ in an unstrained system. Diffusive hops are thermally activated and follow Arrhenius rates $D_{\rm GB}^{(0)} \sim D_0 \exp(-\Delta E_0/kT)$. The activation barrier is reduced under stress by $\Delta E(\sigma) \approx \Delta E_0 - \sigma V^\ast$, where $V^\ast$ is an activation volume of order a unit cell volume. This leads to a diffusion coefficient under stress of $D_{\rm GB}(\sigma) = D_{\rm GB}^{(0)} \exp(\sigma V^\ast/kT)$. Near the yield stress $\sigma \sim 0.1\,\mu$ so perhaps $\sigma V^\ast/kT \sim 1$, justifying $D_{\rm eff} \approx 10^{-6}\, \omega_pa_i^2$ at our assumed grain size and temperature. 
If dislocation nucleation and glide in the bulk contributes comparably at high stress convergence may even be observed at rates as fast as $\dot{\epsilon} \sim 10^{-5}\,\omega_p$.  The volume filling fraction of bulk is an order of magnitude larger and diffusion coefficients for dislocations at finite strain could be faster than vacancies and interstitials (Fig.~\ref{fig:Dall}, dotted curve). Even though their motion depends on similar activation physics dominated by nearest neighbor Coulomb interactions, their formation rates in GBs are likely higher. 

\section{Conclusion}

To reach a total strain of $\varepsilon \sim 0.2$ at a strain rate $\dot{\epsilon} \sim 10^{-6} \,\omega_p$ the required physical time is $t \sim \varepsilon/\dot{\epsilon} \sim 2\times10^{5}\,\omega_p^{-1}$, or roughly $10^{6}$–$10^{7}$ MD timesteps using typical choices of $\Delta t \sim 0.1\text{–}0.01 \, \omega_p^{-1}$. This is computationally feasible on modern GPU-accelerated codes for systems of $N \approx 10^5$ particles ($v_{\rm box}/v_{\rm el} \approx 10^{-3}$), and only requires about an order of magnitude more computing time per simulation than the typical runs used to resolve $D^*_{\rm GB}$ in Fig. \ref{fig:Dall}. Revisiting \cite{HorowitzKadau} using smaller grains and slower strain rates is likely to yield new and interesting results.

\begin{acknowledgments}
This work was supported by a grant from the Simons Foundation (MP-SCMPS-00001470) to MC. This research was supported in part by the National Science Foundation under Grant No. NSF PHY-1748958. Financial support for this publication comes from Cottrell Scholar Award \#CS-CSA-2023-139 sponsored by Research Corporation for Science Advancement. AB is supported by a PCTS fellowship and a Lyman Spitzer Jr. fellowship. 
MC and AB thank the KITP for hospitality and MC acknowledges support as a KITP Scholar. 
\end{acknowledgments}


\begin{thebibliography}{}
\expandafter\ifx\csname natexlab\endcsname\relax\def\natexlab#1{#1}\fi
\providecommand{\url}[1]{\href{#1}{#1}}
\providecommand{\dodoi}[1]{doi:~\href{http://doi.org/#1}{\nolinkurl{#1}}}
\providecommand{\doeprint}[1]{\href{http://ascl.net/#1}{\nolinkurl{http://ascl.net/#1}}}
\providecommand{\doarXiv}[1]{\href{https://arxiv.org/abs/#1}{\nolinkurl{https://arxiv.org/abs/#1}}}

\bibitem[{M.~E. {Caplan} {et~al.}(2022){Caplan}, {Bauer}, \& {Freeman}}]{CaplanBauerFreeman}
{Caplan}, M.~E., {Bauer}, E.~B., \& {Freeman}, I.~F. 2022, \bibinfo{title}{{Accurate diffusion coefficients for dense white dwarf plasma mixtures},} Monthly Notices of the Royal Astronomical Society, 513, L52, \dodoi{10.1093/mnrasl/slac032}

\bibitem[{M.~E. {Caplan} {et~al.}(2025){Caplan}, {Smith}, {Yaacoub}, {Serrano}, {Taira}, \& {Bransgrove}}]{GBD2025}
{Caplan}, M.~E., {Smith}, N.~T., {Yaacoub}, D., {et~al.} 2025, \bibinfo{title}{{Grain Boundary Diffusion in Yukawa Crystals},} arXiv e-prints, arXiv:2510.20980, \dodoi{10.48550/arXiv.2510.20980}

\bibitem[{M.~E. {Caplan} \& D. {Yaacoub}(2025){Caplan} \& {Yaacoub}}]{CaplanYaacoub2025}
{Caplan}, M.~E., \& {Yaacoub}, D. 2025, \bibinfo{title}{{No Evidence of Anomalous Diffusion in Yukawa Crystals},} Research Notes of the American Astronomical Society, 9, 104, \dodoi{10.3847/2515-5172/add1d2}

\bibitem[{C.~J. Horowitz \& K. Kadau(2009)Horowitz \& Kadau}]{HorowitzKadau}
Horowitz, C.~J., \& Kadau, K. 2009, \bibinfo{title}{Breaking Strain of Neutron Star Crust and Gravitational Waves,} Phys. Rev. Lett., 102, 191102, \dodoi{10.1103/PhysRevLett.102.191102}

\bibitem[{J. {Hughto} {et~al.}(2011){Hughto}, {Schneider}, {Horowitz}, \& {Berry}}]{Hughto2011Diffusion}
{Hughto}, J., {Schneider}, A.~S., {Horowitz}, C.~J., \& {Berry}, D.~K. 2011, \bibinfo{title}{{Diffusion in Coulomb crystals},} \pre, 84, 016401, \dodoi{10.1103/PhysRevE.84.016401}


\end{thebibliography}
\end{document}